# The Macroscopic Quantum Effect in Nonlinear Oscillating Systems: a Possible Bridge between Classical and Quantum Physics

Danil Doubochinski and Jonathan Tennenbaum

*Quantix - Société de Recherche et Développement en Technique Vibratoire*
*86, Rue de Wattignies, 75012 Paris, France*
doubochinski@hotmail.com ; tennenbaum@debitel.net

## Abstract

Einstein, de Broglie and others hoped that the schism between classical and quantum physics might one day be overcome by a theory taking into account the essential nonlinearity of elementary physical processes. However, neither their attempts, nor subsequent ones were able to supply a unifying principle that could serve as a starting-point for a coherent understanding of both microphysical and macroscopic phenomena. In the late 1960s the phenomenon of amplitude quantization, or Macroscopic Quantum Effect (MQE), was discovered in a class of nonlinear oscillating systems in which two or more oscillating subsystems are coupled to each other by interactions having a specific phase-dependent character -- so-called argumental interactions. Experimental and theoretical studies of the MQE, carried out up to the present time, suggest the possibility of a new conceptual framework for physics, which would provide a bridge between classical and quantum physics, replacing the Newtonian notion of "force" by a new conception of physical interaction. The present paper provides a brief introduction to the MQE and some ideas about its possible significance in the search for new approaches to the understanding of quantum phenomena.

## 1. Introduction

From the 18[th] century up to and including the whole period of creation of modern quantum and atomic physics, the *simple pendulum* and the *classical linear oscillator* remained the basic models and reference-points for the concept of an "oscillating system". While acknowledging, in a certain way, the *intrinsically oscillatory nature of microphysical objects*, quantum mechanics provided no new physical insight into the nature of oscillating systems as such, and no intelligible explanation for the existence of the quantum of action. Instead, quantization was introduced "ex machina", as a postulate upon which the mathematical structure of quantum mechanics was based.

Einstein, de Broglie and others hoped that the paradoxical nature of quantization and the schism between classical and quantum physics might one day be overcome, by a theory taking into account the *essential nonlinearity* of elementary physical processes. However, neither their attempts, nor subsequent studies of the behavior of systems described by nonlinear differential equations, were able to supply *a new unifying principle*, that could serve as a starting-point for a coherent understanding of both microphysical and macroscopic phenomena.

In 1968 while students of Moscow University, Danil and Yakov Douboshinski discovered the phenomenon of *amplitude quantization* – also known as the *Macroscopic Quantum Effect (MQE)* -- in a large class of nonlinear oscillating systems whose behavior is difficult, if not impossible to understand on the basis of classical theoretical mechanics. The MQE, which can be demonstrated in a variety of electromechanical, electronic, acoustical and other devices, arises when two or more oscillating subsystems are coupled to each other by interactions having a specific phase-dependent character -- so-called argumental interactions. Argumental coupling gives rise to a *new type of oscillating system*, whose properties cannot be reduced to those of the interacting components in any simple way. Ensembles of argumentally interacting macroscopic oscillators typically possess a discrete set of stable quasi-stationary modes and other characteristics strikingly similar to microscopic quantum physical objects. Argumental interactions are impervious to treatment by known mathematical methods of linearization; they require a new conceptual framework, which takes into account the existence of a *general principle of Nature* lying beyond the reach of mathematical analysis per se. This suggests the possibility of creating a bridge between classical and quantum physics; a bridge that would take the form of a unified "Physics of Interaction", in which -- as we shall briefly indicate at the end of this article – the notion of a *true physical interaction* will replace the Newtonian concept of force.

Unfortunately, despite extensive experimental and theoretical investigations of argumental interactions, carried out by the Doubochinskis and their collaborators in the Soviet Union during the 1970s and 1980s especially, the very existence of the MQE remains little-known to the general scientific community today, and its broader implications have not been adequately explored.

This paper is intended to provide a brief introduction to the MQE and the physical principles of argumental interactions, and some preliminary ideas about their possible significance in the search for new approaches to the understanding of microphysical objects such as atoms. Among other things, we present a simple experimental apparatus demonstrating how a discrete, quantized spectrum of stable modes arises in a natural and rather transparent way, through the interaction of two oscillating systems.

We do not pretend that the ideas, presented here, by themselves constitute a replacement for present-day quantum mechanics. Determining the *exact* nature of the relationship between the MQE and the phenomena of quantization in microphysics is an important problem, which the authors intend to address in a forthcoming publication. In the meantime we can affirm, at the very least, that systems based on argumental interactions provide a much richer physical model for the dynamics of interacting oscillatory processes, both macrophysical and microphysical, than has existed until now.

## 2. Doubochinski's Argumental Pendulum

The first demonstration of the MQE was the so-called "argumental pendulum": a pendulum having a discrete set of stable amplitudes of oscillation (Figure 1). As this pendulum is still the best illustration of the principle of argumental interactions, we shall describe it in some detail here. A demonstration apparatus, showing the quantized amplitudes and other main features of this system's behavior, is simple to build and operate, and ought to be standard equipment in physics classrooms.

In its simplest form, the argumental pendulum is composed of two interacting oscillatory processes: (1) a simple pendulum with a natural frequency on the order of 1-2 Hz, with a small permanent magnet fixed at its free end; and (2) a stationary electromagnet (solenoid) positioned under the equilibrium point of the pendulum's trajectory and supplied with alternating current whose frequency can range from tens to thousands of hertz, thus differing very substantially from the pendulum's own proper frequency. The pendulum arm and solenoid are configured in such a way, that the magnet at the end of pendulum arm interacts with the oscillating magnetic field of the solenoid only over a limited portion of its trajectory -- the so-called "zone of interaction" -- outside of which the strength of the magnetic field drops off rapidly to zero. This *spatial inhomogeniety* of the interaction is key to the quantized behavior and other unusual properties of the system.

Setting the pendulum into motion, the following behavior is easy to observe.

1. When released from any given position, the pendulum's motion evolves into a stable, very nearly periodic oscillation, whose amplitude belongs to a *discrete set of possible values* (Figure 2). The existence of a discrete set of stable amplitudes, and a corresponding discrete set of energies of oscillation, suggests a close analogy to the quantized energy-states of atoms and other microscopic quantum systems. It is in this sense that we speak of a *"Macroscopic Quantum Effect (MQE)"*, of *"quantized modes"* and *"quantized amplitudes"* in the present article.

2. The stability of each amplitude-mode is maintained by a constant self-adjustment of the phase relationship between the pendulum and the high-frequency field. Through its interaction with the field, the pendulum extracts the amount of energy needed to compensate its frictional losses for a given period.

3. The values of the quantized amplitudes -- and the corresponding energies of the quantized modes -- are essentially independent of the strength of the alternating current supplied to the electromagnet, over a very large range. The pendulum compensates for the changes in the strength of the magnetic field, by slightly shifting the phase of its entry into the zone of interaction, while maintaining almost exactly the same amplitude and frequency. If we gradually reduce the strength of the current in the electromagnet, we reach a threshhold below which the stable mode can no longer support itself, and the pendulum decays into a lower-energy mode.

4. The number of available stable modes and the values of the corresponding amplitudes, depend strongly on the frequency of the alternating current in the electromagnet. The higher the frequency, the larger the number of stable modes that can be excited.

5. The maintenance of quantized amplitudes in the argumental pendulum is connected with a mechanism of exchange of energy, which differs fundamentally from the familiar classical case of "forced oscillations" of an oscillator under the action of a periodic external force. In the classical case, an efficient exchange of energy between the oscillator and the external signal occurs *only* when the frequency of the external force is close to the proper frequency of the oscillator. This is the classical reference-point for the phenomenon of resonance. In the case of the argumental pendulum, in contrast, stable oscillations are maintained by an efficient coupling between subsystems whose frequencies can differ by *two or more orders of magnitude*. The pendulum oscillates in a stable mode at frequencies close to its own proper (undisturbed) frequency.

As we mentioned, the amplitude quantization and other novel characteristics of the argumental pendulum, absent from the classical theory of oscillations, originate in the specific character of the interaction between the pendulum and the electromagnet. Here, in contrast to the systems commonly studied in classical mechanics, the exchange of energy is regulated mainly by the *phase relationships* of the interacting periodic processes – more specifically, by the relationship of phases of the field at the moment the pendulum enters and exits the zone of interaction. Within that zone, the oscillating magnetic field causes the pendulum to experience a series of alternately accelerating and decelerating impulses, depending on the momentary polarity of the current. That train of impulses is interrupted at the moment the pendulum leaves the zone. If the *transit time*, taken by the pendulum to traverse the interaction zone, corresponds to an *integral number of periods of the electromagnet*, then the effects of the accelerating and decelerating half-cycles will cancel out, and the net effect will be zero[1]. But if the pendulum leaves the zone of interaction after a *non-integral* number of periods of the electromagnet, relative to the moment of entry into the zone, then a net transfer of energy will occur, and the pendulum will experience a net non-zero accelerating or decelerating effect from its interaction with the alternating field. By varying its phase relative to the electromagnet the pendulum is able, in a sense, to *self-regulate* its interaction with the electromagnet, making possible stable motions in which the frictional losses are compensated by the energy extracted from the magnetic field over a given period of the pendulum's oscillation.

These observations permit us to obtain a first, preliminary glimpse of the origin of the quantized amplitudes: A *necessary* condition for maintenance of a quasi-stationary phase relationship, is that the *time* between successive entries of the pendulum into the interaction zone, be equal to an *integral multiple* of the period of the electromagnet. That series of values obviously forms a *discrete set*. The time between successive entries, on the other hand, is very nearly[2] equal to a half-period of the pendulum, which in turn depends upon its amplitude according a well-known formula from the classical theory of the pendulum[3]. Consequently, stable quasi-stationary motions can occur only for a discrete set of values of the amplitude.

We hasten to point out, that the preceeding reasoning does not explain why the quasi-stationary modes, observed experimentally, *actually exist*; nor does it account for their stability, which depends upon a complex interplay between fluctuating phases and amplitudes. An exhaustive treatment of this problem can be found in the publications [1-4, 6, 7, 9, 11] For the moment, we want to call special attention to the following features of the argumental pendulum, which typify the general case of oscillating systems based on argumental interactions:

While interacting, the two subsystems (the pendulum and the electromagnet) retain their integrity as oscillating systems, each operating at or near its proper frequency and "living", so to speak, in its own proper time.

The self-regulating character of the interaction gives rise to a discrete array of stable modes of existence of the interacting ensemble. In each of those stable modes the subsystems are bound together in a *dynamic unity* -- a third physical object in its own right -- characterized by its own specific oscillatory parameters. This new object maintains itself on the basis of *phase fluctuations*: small retardations and advances in the phase of the pendulum relative to the alternating magnetic field, which regulate the magnitude and direction of the momentary flow of energy between the two subsystems, and thereby insure the stability of the given mode.

The physical content of the interaction lies not only in a constant exchange of energy between the interacting subsystens, but also in the continual *transformation of the form of energy* (action) -- in this case, between gravitational, mechanical (kinetic), magnetic and electric. Here it is important to consider not only the accelerating or decelerating influence of the electromagnet on the pendulum, but also the inductive effect of the pendulum's motion on the current in the electromagnet, and indirectly on the current source, which we have not included explicitly in our description of the system. This reverse coupling plays a significant role in the case of multiple pendula interacting with a common electromagnet (see next section).

In overall effect, the argumental mechanism accomplishes a transformation of energy between oscillatory systems operating at widely differing frequencies, and whose physical natures can be different. This leads to a variety of interesting technological applications, discussed elsewhere.[4] The argumental principle might also explain how the transformation of energy occurs in many natural processes, not adequately understood until now.

Before examining the main ideas of a general theory of argumental interactions, it is worthwhile to briefly present two further examples, of particular interest for atomic and quantum physics.

## 3. Multiple argumental pendula

We can look at the single argumental pendulum, just described, as very roughly analogous to a *single-electron atom*. Here the "electron" corresponds to the pendulum and the nucleus to the high-frequency oscillating magnetic field, the two being bound together by a phase-regulated (argumental) exchange of energy. It is easy, however, to extend this to a "zeroth approximation" model of a *multi-electron atom*. The mechanism of argumental interactions makes its possible for a single high-frequency source to support not just one, by an *entire ensemble* of argumental oscillators, simultaneously, each operating at a frequency near its own proper frequency (which can be the same or different from the others) and each with its own discrete array of stable amplitudes.

A simple modification of the experimental setup described above, provides an impressive demonstration of this case: a series of pendula, of different lengths, oscillate in parallel vertical planes, each interacting with the same elongated solenoid and each at a frequency near its own proper frequency. While the pendula appear to oscillate independently of each other, their motions are in fact correlated, both through the direct interplay of their magnetic fields as well as via their inductive interaction with the common electromagnet. Preliminary investigations[5] pointed to novel features of the behavior of multiply-coupled argumental oscillators, that may be relevant to atomic and quantum physics.

## 4. The Argumental Analog of Planck's Elementary Oscillator

Another case worth mentioning here is an analog of the hypothetical "elementary oscillator" which Max Planck employed in his original analysis of the interaction of matter and radiation. This case, which has considerable *theoretical* interest but is difficult to realize experimentally in a macroscopic system, was studied using mathematical analysis and computational methods [8]. Consider a mass, fixed at the end of a spring and free to oscillate in the x-direction, and consider the interaction of this oscillator with an electromagnetic wave propagating in the x-direction. The classical differential equation for this system has the form

$$m (\ddot{X} + 2\beta \dot{X} + \omega_0^2 X ) = E\, q\, \sin(\Omega t - kX) \qquad (1)$$

Here **X** denotes the coordinate of the position, **m** denotes the mass, **q** the charge it carries, and **E** the longitudinal component of the wave, evoked by its interaction with the charge. The fundamental distinction between this system and the classical "elementary oscillator" is the dependence of the force, represented by the right side of the equation, on the *position* of the oscillating charge, rather than merely the time. As in the case of the argumental pendulum, but in a slightly different way, this spatial dependence permits the oscillator to modulate its interaction with the field via changes in the phase relations. Investigations showed that when the wave frequency $\Omega$ is large compared to the proper frequency $\omega_0$ of the oscillator, this self-modulated interaction of the oscillating charge with the wave gives rise to a discrete series of stable amplitudes of the oscillator -- without any need for additional assumptions.

The possibility of this sort of effect was apparently overlooked by the founders of quantum theory at the beginning of the 20th century. It is reasonable to suspect, that quantum physics would have developed in a significantly different direction, if the results, described here, had been known to Planck, Einstein, Sommerfeld, Heisenberg, Schrödinger and others in that period.

This does not mean to imply that the MQE suffices, in and of itself, to explain Planck's quantum of action and the phenomena described by Schrödinger's wave function. We are presently investigating the relationship between them. The example of the MQE does suggest, among other things, that the mechanism of quantization should be sought more in the *processes of interaction* of microphysical objects, than in the objects per se. Such an approach would be closer to the original standpoint of Max Planck, than the later (1905) suggestion by Einstein, according to which electromagnetic radiation should be regarded as already quantized, independently of its interaction with material systems.[6]

Studies of the argumental interaction between an oscillator and wave revealed the interesting fact, that the effect of quantization of amplitudes does not depend upon the presence of dissipation, but persists even when the coefficient β, characterizing the dissipative energy loss, is equal to zero. *Generally speaking, what is essential to the genesis of discrete ("quantized") amplitudes, is the constant, phase-modulated mutual exchange of energy between the interacting oscillatory systems -- a dialog organized via the argumental interaction, without the need for dissipation and without the participating systems themselves having to be explicitly nonlinear.*

## 5. The Theory of Argumental Oscillations

The theory of argumental oscillations was laid out in a series of publications in Soviet journals during the 1970s and 1980s, and presented at international symposia. Its development was based on extensive experimental investigations and computer calculations. These investigations made it possible to understand the main qualitative features of argumental oscillations, and to develop practical algorithms for calculating the values of the quantized amplitudes for the pendulum and other systems operating on the argumental principle, under various conditions.

The theory of argumental oscillations is nearly unique in the domain of oscillatory physics, insofar as it does not depend on the ad hoc linearization methods which – for lack of a better alternative! – still constitute the basic tool in most mathematical treatments of nonlinear systems. Furthermore, the relationship of the theory of argumental interactions to classical physics is *paradoxical*. At first glance, the theory of argumental oscillations does not directly contradict the laws of classical physics, and might appear merely to be a new branch of the classical theory of oscillations. On the other hand, it presupposes a subtle kind of "phase intelligence" on the part of real physical systems, which has no obvious classical explanation. That "intelligence" permits the system to steer its evolution into one of a discrete set of stable regimes and to maintain its stability by a process of continual phase fluctuations. Experimental studies have shown, for example, that the motion of the argumental pendulum in its discrete regimes is *never strictly periodic*. Instead, the pendulum's space-time trajectory wanders from one cycle to the next, slightly shifting its phase relative to that of the magnetic field of the solenoid and varying its amplitude within a small "exploratory" interval around one of the stable values. These fluctuations allow the oscillator to "feel" its environment and react to it. Perhaps it is not too far-fetched to suggest, that the MQE depends upon a dynamic characteristic of real physical systems, lying in the boundary-area between nonliving and living processes.

It is indispensible to point out, that the discrete stable amplitudes, observed in actual physical experiments, do not manifest themselves in the usual sorts of computer modelling of the nonlinear differential equations describing argumental interactions. In order to successfully model the discrete sets of stable amplitudes of the argumental pendulum and similar systems, special computational codes were developed, taking into account the qualitative features of the MQE.

To grasp the main ideas of the general theory of argumental oscillations, it is useful to examine the case of the argumental pendulum once again, from a somewhat different point of view than we have taken so far.

Imagine first the case, in which the pendulum has no initial velocity, but hangs in the equilibrium position inside the interaction zone. In this case, under the influence of the alternating magnetic field of the solenoid the pendulum executes small forced oscillations around the equilibrium point, in accordance with the classical theory. Since the system is very far from resonance conditions, the pendulum's amplitude remains small, and it is obliged to oscillate at the frequency of the magnetic field, with practically no trace of its own proper frequency. In this state the pendulum is virtually the *"slave"* of the external force.

Now imagine the case, where we give the pendulum a initial impulse sufficient to carry it well beyond the interaction zone. Now the pendulum is in a sense *its own master*: its gross motion is dominated by its own oscillatory characteristics, oscillating with a frequency close to its proper frequency, but whose amplitude and phase undergo constant variations as a result of the interaction with the electromagnet. A net acceleration or net deceleration of the pendulum, as the result of a single transit through the interaction zone, causes a shift (retardation or advancement) in the phase of the (very nearly) periodic function expressing the angular *position* of the pendulum, relative to the time-function of its angular *velocity*. This representation of the system evokes significant analogies to the analysis of phase- and frequency-modulated signals in radio theory.

Consider, next, the time-function of the field experienced by the pendulum in the course of its nearly-periodic motion. It is easy to see that it consists of a sequence of short, high-frequency "bursts", each corresponding to a transit of the pendulum through the interaction zone, and recurring at intervals of a half-period of the pendulum. This signal contains a rich spectrum of frequency components. Under certain conditions, in fact, that spectrum will include a component whose frequency is close to the pendulum's own proper frequency. This opens up the possibility, that the pendulum might be able, in a sense, to *resonate with one of the spectral components generated by its own modulation of the original signal*, and to thereby draw net energy from the field.

One can now easily anticipate the first steps of mathematical analysis, showing quite clearly how the condition of self-resonance might arise. Letting **X** signify the pendulum's angle relative to the equilibrium position, the motion of the pendulum can be described by a differential equation of the form:

$$m (\ddot{X} + 2\beta \dot{X} + \omega_0^2 X ) = A f(X) \sin(\Omega t). \qquad (2)$$

Here **sin(Ωt)** is the periodic function of the magnetic field, **A** its overall amplitude and **f(X)** is a function specifying the relative strength of the magnetic field at a given angular position of the pendulum. The experimental apparatus is constructed in such a way, that the value of **f(X)** is close to 1 when the pendulum is inside the interaction zone ( **-x₀ < x < x₀**), and drops off rapidly to zero for values of **x** outside that interval. (The following discussion will make it plausible, however, that the amplitude quantization effect itself does not depend on the exact form of the function **f**, but only the numerical values of the stable amplitudes. Carefully analysis confirms this conclusion.)

Needless to say, equation (**2**) cannot be solved mathematically in closed form. Assume, in accordance with what we have already said, that the pendulum's motion can be represented, in first approximation, by a simple periodic function

$$X = a \sin(\omega t),$$

where $\omega$ is close to the proper frequency of the pendulum. In this case **f(X)** is a periodic function having the same period as x, and can be expanded in a Fourier series:

$$f(X) = f(a \sin(\omega t)) = \sum_{n=1}^{\infty} C_n(a) \sin(n\omega t). \qquad (3)$$

The dependence of the Fourier coefficients on the pendulum's amplitude **a**, plays an essential role in the genesis of the quantized amplitudes, because it creates a *"feed-back" between the amplitude and the distribution of the frequency components of the signal experienced by the pendulum*. Taking account now of the high-frequency periodic character of the magnetic field, expressed in the factor $\sin(\Omega t)$, we can now expand the right-hand side of equation (**2**) – i.e. the time-function of the field action experienced by the pendulum --, in the form:

$$A f(X) \sin(\Omega t) = A \sum_{n=1}^{\infty} C_n(a) \sin(n\omega t) \sin(\Omega t) =$$

$$= A \sum_{n=1}^{\infty} C_n(a) \tfrac{1}{2} [\cos(\Omega - n\omega)t - \cos(\Omega + n\omega)t] \qquad (4)$$

In this way, the space-time modulation of the oscillating magnetic field by the pendulum's motion, generates a spectrum of frequencies extending upward and downward from $\Omega$ in steps of $\omega$. Assuming now that $\Omega$ is an integral multiple of $\omega$, i.e. $\Omega = N\omega$ for some integer **N**, the spectrum (**4**) will contain the term $\tfrac{1}{2} A C_{N-1}(a) \cos(\Omega - (N-1)\omega)t$ whose frequency coincides with that of the pendulum. This points to the possibility of a resonant interaction between the pendulum and this spectral component, generated by the pendulum's own modulation of the magnetic field.

At this point we can begin to see how a discrete series of stable amplitudes might possibly arise. If it were possible to simply ignore all the periodic components other than $\sin(\omega t)$ on the left and right sides of equation (**2**), then comparing the two sides would lead to a second relationship between the amplitude **a** and the value of the corresponding Fourier coefficient, $C_{N-1}(a)$. Since the coefficient $C_{N-1}(a)$ oscillates strongly with increasing values of **a** -- due to the sharp inclines in the profile of the function f -- the resulting equation in **a** has a *discrete series of solutions*. Such a discrete set of solutions will exist, in general, for each value of the integer **N**, for which $\Omega/N$ lies sufficiently close to the proper frequency of the pendulum.

In reality, of course, the higher-frequency components cannot simply be ignored, and the motion of the pendulum differs significantly from a simple harmonic motion. As we already noted above, small retardations and advances in the phase of the pendulum, relative to a simple harmonic motion, are essential to the mechanism of the MQE; it is by this means that the pendulum is able to precisely regulate its exchange of energy with the magnetic field, in such a way as to compensate frictional losses and maintain a quasi-stationary mode.

Accordingly, the mathematical theory of argumental oscillations begins by assuming instead of **X = a sin(ωt)** a pendulum motion of the form

$$X = a(t) \sin(\omega t + \phi(t)),$$

where the amplitude **a(t)** and phase shift **ϕ(t)** are arbitrary, slowly-varying functions of the time. This leads to a more complicated set of differential equations, which can be analyzed using the averaging method of Krylov-Bogoliubov. The result is a method which not only allows the quantized amplitudes to be calculated, but also establishes the stability of the corresponding modes.

We cannot go more into the details of these investigations here, but wish merely to emphasize a very important conclusion: The principles of argumental interactions, elaborated above with reference to the argumental pendulum as an exemplary case, apply in fact to a very large class of oscillatory systems. In this context, the function defining the spatial dependence of the interaction – f(x) in our discussion above -- plays a role strikingly analogous to that of the "potential well" in quantum mechanics.

## 6. Toward a physics of interaction

How could the MQE and the principle of argumental interactions, be of use in developing a deeper and more correct understanding of microphysical processes, including atomic structure? At this point we are very far from being able to offer an elaborated alternative theory, but we would like to make the following remarks in closing:

It is reasonable to expect, that the Schrödinger wave function is merely an envelope, average or asymptotic limit, expressing the outcome of interactions that are essentially argumental in nature and occur on much smaller temporal and spatial scales, than the period and wavelength of the wave function itself. We believe that one of the main difficulties, inhibiting progress in this domain, has been the lack of macroscopically accessible examples of oscillatory systems whose behavior differs fundamentally from that of the well-known classical cases.

The principle of argumental interactions opens up a qualitatively different, physically much richer class of potential experimental models for the interaction of oscillating systems, than have been used until this time. If the possibilities of present-day technology were fully exploited in creating a new collection of experiment models for ensembles of argumentally interacting oscillators, this would doubtless produce a great wealth of new phenomena, relevant to our understanding of atomic and nuclear processes.

At present the most important task is to elaborate a *new, precise conception* of the *nature of physical interactions,* radically different from the one that has prevailed in the teaching of physics since the time of Newton and Descartes, and which might provide a qualitatively superior insight into the nature of so-called elementary particles, and of compound systems such as nucleii, atoms and molecules, than is possible on the basis of presently-existing notions. The new conception would relegate "forces" to the status of mere *effects* of the interactions between physical systems. Taking the case of argumental interactions and the MQE as a concrete point of reference, we propose to develop such a conception along the following summary lines:

1. Real physical systems are fundamentally oscillatory in nature. Each physical system lives in its own time, characterized first of all by its own proper frequency or period, and by a phase variable which corresponds to the succession of physical states (transformations) constituting a single cycle of action of the system.

2. Real physical interactions are fundamentally oscillatory in nature. Each such interaction involves a continual process of transformation between different forms of energy (i.e. different forms of organization of physical action).

3. A real interaction between physical systems presupposes the existence of a connection or coupling between them, accompanied by a self-regulated exchange of energy, by virtue of which each component system undergoes a certain continual internal modification (modulation), while at the same time retaining its essential integrity and continuing to "live" in its own proper time.

4. An ensemble of systems interacting in this way, constitutes a new physical entity – a dynamic unity -- possessing its own specific oscillatory and other physical characteristics, which in general are completely different from those of the components from which it was formed.

5. The connections or "bonds" of interaction, formed between oscillating systems, should be regarded as real physical objects in their own right.

We shall present our ideas on these matters in more systematic fashion in a forthcoming paper.

**Footnotes**

1. This reasoning assumes that the intensity of the magnetic field is essentially constant as a function of position within the interaction zone, and drops instantly to zero at the boundary of the zone. These conditions are realized only in rough approximation in the demonstration version of the argumental pendulum. It turns out, however, that the Macroscopic Quantum Effect does not depend on the exact form of spatial dependence of the field strength, as long as it is strongly inhomogeneous. See Section 4 of this paper.

2. In actuality, the time between successive entries of the pendulum into the interaction zone varies slightly as a result of variations in the transit time across the interaction zone. The transit time in turn depends in a complicated way on the phase of the field and the velocity of the pendulum at the moment of entry into the zone.

3. The dependence of frequency $\omega$ upon amplitude $a$ in the simple pendulum can be represented by the approximate formula: $\omega = \omega_0 (1 - \mu a^2)^{1/2}$, where $\omega_0$ is the proper frequency of the pendulum (frequency for very small amplitudes), and $\mu$ is a constant depending on the construction of the pendulum.

4. Potential technological applications of argumental oscillations include new types of motors and electrical propulsion and generating devices, systems for energy conversion, methods for the mixing and separation of different physical media, desalination and adiabatic refrigeration. A draft article, devoted to the last of the mentioned applications – presently the closest to commercialization – is available on request from the authors.

5. Experiments and theoretical investigations of multiply-coupled oscillators, including the behavior of such oscillators in a common radiation field, were carried out in the 1970s by a special laboratory directed by Danil Doubochinski at the Vladimir State Pedagogical Institute.

6. We are referring to Einstein's famous paper**,** "On a Heuristic Viewpoint Concerning the Production and Transformation of Light" [10] in which he states, among other things: "Energy, during the propagation of a ray of light, is not continuously distributed over steadily increasing spaces, but it consists of a finite number of energy quanta localised at points in space, moving without dividing and capable of being absorbed or generated only as entities."

## Literature

1. D.B. DOUBOCHINSKI, Ya.B. DUBOSHINSKY at al., *Oscillations with self-regulating interaction time*. Dokl. Akad. Nauk SSSR, 204, N°5, 1065 (1972). [Sov. Phys. Doklady 17, 541(1972)].

2. D.B. DOUBOCHINSKI, Ya.B. DUBOSHINSKY et al., *Asynchronous excitation of undamped oscillations*. Usp. Fiz. Nauk 109, 402 (1973) [Sov. Phys.-Usp. Vol 16, 158 (1973)].

3. D.B. DOUBOCHINSKI, D.I., PENNER, Ya.B. DUBOSHINSKY, *Über die Erregung mechanischer asynchroner Schwingungen*. Experimentelle Technik der Physik, XXI, 5, Berlin, 1973, pp. 397- 401.

4. D.B. DOUBOCHINSKI, Ya.B. DUBOSHINSKY and D.I. PENNER, *Argumental oscillations*. Izvestia AN SSSR, Mechanics of Rigid Bodies, N° 1,  18, (1975) [Sov. Phys. Izv. 7, 54 (1975)].

5. D.B. DOUBOCHINSKI, Ya.B. DOUBOCHINSKY et al. *Electromagnetic model of the interaction of resonators*. Dokl. Akad. Nauk SSSR, 227, N°3, 596 (1976) [Sov.Phys.Doklady 27, 51(1976)].

6. Ya.B. DUBOSHINSKI, L.A. VAINSHTEIN*, Excitation of low-frequency oscillations by a high-frequency force*. Zh. Tech. Fiz. 48, 1321 (1978) [Sov. Phys.-Tech. Phys. 23, 745 (1979)].

7. D.B. DOUBOCHINSKI, Ya.B. DUBOSHINSKY et al., *Discrete modes of a system subject to an inhomogeneous  high-frequency force*. Zh. Tech. Fiz. 49, 1160 (1979) [Sov. Phys.-Tech. Phys. 24, 642 (1979)].

8. D.B. DOUBOCHINSKI, Ya.B.  DUBOSHINSKY, *Wave excitation of an oscillator having a discrete series of stable amplitudes*. Dokl. Akad. Nauk SSSR, 265, N°3, 605 (1982) [Sov. Phys. Doklady 27, 564 (1982)].

9. D.B. DOUBOCHINSKI, Ya.B. DUBOSHINSKY, *Amorcage argumentaire d'oscillations entretenues avec une série discrète d'amplitudes stables -- Argumental Excitation of Continuous Oscillations with Discrete Set of Stable Amplitudes*. E.D.F.-Bulletin de la Direction des Etudes et Recherches, Serie C, N°1, 1991.

10. A. EINSTEIN, *Über einen die Erzeugung und Verwandlung des Lichtes betreffenden heuristischen Gesichtspunkt*. Annalen der Physik 17, 132 (1905).

11. J. TENNENBAUM  *Amplitude Quantization as an Elementary Property of Macroscopic Vibrating Systems*. 21st CENTURY SCIENCE & TECHNOLOGY - USA Winter 2005 – 2006.

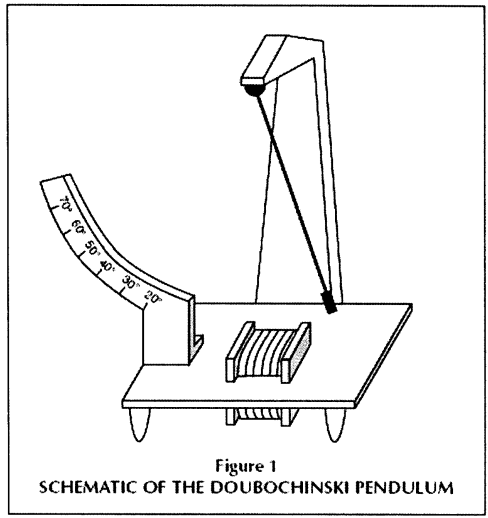

**Figure 1**
SCHEMATIC OF THE DOUBOCHINSKI PENDULUM

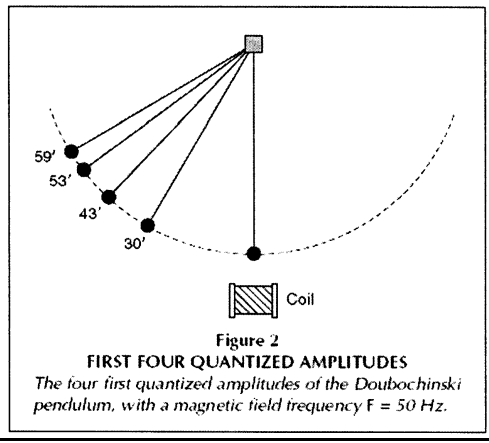

**Figure 2**
**FIRST FOUR QUANTIZED AMPLITUDES**
The four first quantized amplitudes of the Doubochinski pendulum, with a magnetic field frequency F = 50 Hz.